\documentclass[twocolumn,superscriptaddress,amsmath,amssymb,floatfix,aps,notitlepage,nokeys,pra,longbibliography]{revtex4-1}
\usepackage{txfonts}
\usepackage{amssymb}
\usepackage{mathrsfs}
\usepackage{graphicx}
\usepackage{dcolumn}
\usepackage{bm}
\usepackage{braket}
\usepackage{epsfig}
\usepackage{color}
\usepackage{ulem}
\usepackage{bibunits}
\usepackage{inputenc}
\usepackage[T1]{fontenc}
\usepackage[bb=dsserif]{mathalpha}
\graphicspath{ {./images/} }
\usepackage{hyperref}
\hypersetup{nolinks=true}
\begin{document}
\title{Nearly flat Chern band in periodically strained monolayer and bilayer graphene}

\author{Xiaohan Wan}
\affiliation{%
Theoretical Division, T-4 and CNLS, Los Alamos National Laboratory, Los Alamos, New Mexico 87545, USA
}
\affiliation{%
Department of Physics, University of Michigan, Ann Arbor, MI 48109, USA
}

\author{Siddhartha Sarkar}
\affiliation{%
Department of Physics, University of Michigan, Ann Arbor, MI 48109, USA
}

\author{Kai Sun}
\email{sunkai@umich.edu}
\affiliation{%
Department of Physics, University of Michigan, Ann Arbor, MI 48109, USA
}

\author{Shi-Zeng Lin}
\email{szl@lanl.gov}
\affiliation{%
Theoretical Division, T-4 and CNLS, Los Alamos National Laboratory, Los Alamos, New Mexico 87545, USA
}
\affiliation{%
Center for Integrated Nanotechnologies (CINT), Los Alamos National Laboratory, Los Alamos, New Mexico 87545, USA
}

\begin{abstract}
The flat band is a key ingredient for the realization of interesting quantum states for novel functionalities. In this work, we investigate the conditions for the flat band in both monolayer and bilayer graphene under periodic strain. We find topological nearly flat bands with homogeneous distribution of Berry curvature in both systems. The quantum metric of the nearly flat band closely resembles that for Landau levels. For monolayer graphene, the strain field can be regarded as an effective gauge field, while for Bernal-stacked (AB-stacked) bilayer graphene, its role is beyond the description of gauge field. We also provide an understanding of the origin of the nearly flat band in monolayer graphene in terms of the Jackiw-Rebbi model for Dirac fermions with sign-changing mass. Our work suggests strained graphene as a promising platform for strongly correlated quantum states. 

\end{abstract}
\date{\today}
\maketitle

\section{Introduction}

Electron correlation is at the center of condensed matter physics research and is the driving mechanism for a variety of novel quantum states of matter. There are several known routes to tune the system into a strongly correlated region. In the Landau Fermi liquid, the Coulomb interaction can dominate the electron kinetic energy by tuning the density of the electron to the dilute region. The strongly correlated region can also be achieved using localized electron orbitals, such as electrons in the $d$ shell or $f$ shell. This is illustrated in two paradigmatic models in condensed matter physics: the Hubbard model and the Anderson lattice model for $f$ electrons. In these correlated systems, the physics connects to the atomic limit adiabatically. For instance, by using more and more localized electron orbitals, the overlap of the orbital wave function between neighboring sites can be progressively reduced to approach the atomic limit. Therefore, the narrow electron band in these systems generally has a trivial band topology. 
Another route is to take advantage of the interference effect of the electron wavefunction when electrons hop in the lattice. For certain lattices, such as the Lieb lattice and Kagome lattice, there exists an exactly flat band. The condition for the existence of exactly flat band was derived by Lieb~\cite{PhysRevLett.62.1201}. Nearly flat bands have been achieved in carefully designed Hamiltonians~\cite{PhysRevLett.106.236803,PhysRevLett.106.236802,PhysRevLett.106.236804}. Recently, moir\'e superlattices have emerged as an exciting platform for attaining a nearly flat band through the interference effect of the electron wavefunction~\cite{bistritzer2011moire}. Moir\'e superlattice appears when two incommensurate periodic structures are superposed together. This can be achieved by twisting a two monolayer material with respect to each other or by placing a layer of two-dimensional material on top of another material with a different lattice parameter. As a result of incommensuration, a superlattice with much longer lattice periodicity arises. (Rigorously speaking, the superlattice is in general not periodic. But at the low-energy scale, the incommensuration effect is not important, and the system can be treated as a periodic lattice.) Here, we emphasize that the narrow band in the moir\'e superlattice is not merely due to the reduced energy scale for a small Brillouin zone in the superlattice with a long period. This can be seen by checking the ratio between the band gap and the bandwidth, which can be large under certain conditions. The single particle band is generally not exactly flat in a moir\'e superlattice~\cite{bistritzer2011moire}, except for certain special parameters~\cite{tarnopolsky2019origins}. Nevertheless, these nearly flat bands are sufficient to stabilize a plethora of correlated states, such as superconductivity, correlated insulating states, and Wigner crystals~\cite{cao2018correlated,cao2018unconventional,lu2019superconductors,yankowitz2019tuning,polshyn2019large,xie2019spectroscopic,kerelsky2019maximized,cao2020strange,jiang2019charge,choi2019electronic,zondiner2020cascade,wong2020cascade,nuckolls2020strongly,he2021symmetry,liu2020tunable,regan2020mott,wang2020correlated,xie2020nature,wu2020collective,su2020current,padhi2018doped,padhi2021generalized,padhi2019pressure,stefanidis2020excitonic,bultinck2020mechanism,Xu_Liu_Mak_Shan_2020}.        

The third route is to apply a magnetic field which quenches the kinetic energy of electrons in the plane perpendicular to the magnetic field. Electrons occupy the Landau levels, which are excatly flat in the clean bulk. Moreover, the system is topological as evidenced by an integer quantized Hall conductance. The Landau levels have other unique properties, such as saturation of the trace condition. These properties make Landau-level systems important for stabilizing exotic quantum states, such as the fracitonal quantum Hall state and many other states with nontrivial topological order.

% It is natural to seek similar Landau level physics in material but without an external magnetic field. 
It is natural to seek similar Landau level physics without an external magnetic field. This is indeed possible in systems where strain can be regarded as a pseudo magnetic field. This idea has been explored extensively in the context of graphene~\cite{Vozmediano_Katsnelson_Guinea_2010}, and the formation of Landau levels has also been observed in experiments~\cite{Levy_Burke_Meaker_Panlasigui_Zettl_Guinea_Neto_Crommie_2010}. Unlike the physical magnetic field, the strain does not break the time-reversal symmetry. This means that the pseudo-magnetic field must be opposite for Dirac fermions in two different valleys in graphene. The pseudo magnetic field can also stabilize nontrivial topological phase in graphene as was recently demonstrated in Ref.~\cite{PhysRevLett.128.176406}. Experimentally, the narrow band in graphene under periodic strain was reported~\cite{Mao_Geim_2020} and was investigated theoretically in Refs.~\cite{PhysRevB.102.245427,gao2023untwisting,mahmud2023percolation}. Therefore, strain engineering represents an important direction to achieve flat bands, parallel to the efforts to identify materials with intrinsic flat bands~\cite{Regnault_Xu_Li2022}. 

In this work, we study the occurrence and condition of a nearly flat band in both monolayer and bilayer graphene under periodic strain. We identify nearly flat bands with a nonzero valley resolved Chern number in monolayer graphene. This flat band can be understood based on the Jackiw-Rebbi zero mode in 1+1 dimension Dirac fermion with spatially varying mass. We then go beyond the simple Dirac fermion with winding number 1 to Dirac fermion with winding number 2 stabilized in bilayer graphene. For Dirac fermions in graphene, the strain field can only shift the position of the Dirac cone before the Dirac cone annihilates with the other Dirac fermion in the opposite valley. In this case, the strain can be regarded as a pseudo magnetic field. In bilayer graphene, in addition to the shift of the position of the Dirac cone, the Dirac fermion with winding number 2 can split into two Dirac fermion with winding number 1 or other combinations such as 3 Dirac cone with winding 1 and one cone with winding -1. In this case, the strain field decomposes into two separate sectors: symmetric and antisymmetric (under inversion). In analogy to strain in single layer systems, the symmetric sector results in an effective gauge field. In contrast, the antisymmetric sector provides a distinctive coupling between strain and electronic degrees of freedom, which can no longer be treated as an effective gauge field, and results in a new family of moire systems. We will show the appearance of nearly flat bands with almost ideal quantum geometry as Landau levels in monolayer and bilayer graphene. 

The remainder of the paper is organized as follows. In Sec. II, we review the treatment of strain as a pseudo gauge field in monolayer graphene. In Sec. III, we present results of nearly flat band in monolayer graphene. In Sec. IV, we turn to the flat band in periodic strained bilayer graphene. In Sec. V, we provide an understanding of the occurrence of flat bands in strained monolayer graphene from the perspective of the Jackiw-Rebbi model. We end the paper with a brief discussion and conclusion in Sec. VI.

\section{Strain}
\begin{figure}
\centering
\includegraphics[width=\linewidth]{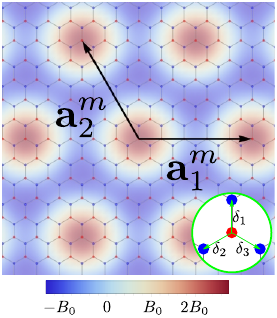}
\caption{Schematic view of periodically strained graphene. $\mathbf{a}^{m}_1=\sqrt{3}N_n a(1,0)$ and $\mathbf{a}^{m}_2=\sqrt{3}N_n a(-\frac{1}{2},\frac{\sqrt{3}}{2})$ are the superlattice lattice vectors. The color map is of the pseudo magnetic field defined in Eq.~\eqref{eq:B field}.
} 
\label{fig:f1}
\end{figure}
Here we briefly review how the strain field can be modeled as a pseudo magnetic field in graphene.
The valley projected Hamiltonian of monolayer graphene around $K_{+}$ is:
\begin{equation}
H_{K_{+}}=\frac{3}{2} a t
\begin{pmatrix}
 0 & q_{x}+i q_{y} \\
 q_{x}-i q_{y} & 0 
\end{pmatrix} 
= \frac{3}{2} a t\begin{pmatrix}
 0 & -2i\partial_{\bar{z}} \\
 -2i\partial_z & 0 
\end{pmatrix},
\end{equation}
where $\partial_{z} = \frac{1}{2}(\partial_x - i\partial_y)$, $\partial_{\bar{z}} = \frac{1}{2}(\partial_x + i\partial_y)$, $a$ is the bond length of graphene, $t$ is the nearest-neighbor hopping strength. 
After we apply a deformation field $\vec{u}$ to monolayer graphene, the hopping strength along $\vec{\delta}_1$,  $\vec{\delta}_2$ and  $\vec{\delta}_3$ (as shown in Fig.~\ref{fig:f1})are modulated by the strain field. Denote the hopping strength along $\vec{\delta}_1$, $\vec{\delta}_2$ and $\vec{\delta}_3$ as $t_1$, $t_2$ and $t_3$ respectively. Then the Hamiltonian is modified as (we use $\hbar=e=c=1$):
\begin{equation}
H_{K_{+}}=\frac{3}{2} a t \begin{pmatrix}
 0 & -2i\partial_{\bar{z}}+e (A_{x}+iA_{y})  \\
-2i\partial_{z}+e(A_{x}-iA_{y}) & 0 
\end{pmatrix},
\end{equation} 
where $e v_{F}A_{x}=\frac{1}{2}(\delta t_{2}+\delta t_{3}-2\delta t_{1}),\,e v_{F} A_{y}=\frac{\sqrt{3}}{2}(\delta t_{3}-\delta t_{2})$ and $v_{F}=\frac{3}{2} a t$. 
The relationship between $\delta t_i$ and the strain tensor $u_{ij}$ is:
\begin{equation}
\begin{split}
t_i &=t+\frac{\partial t}{\partial a} \delta a\\
&=t+\frac{\partial \ln{t}}{\partial \ln{a}}\frac{t}{a} \hat{\delta_i}\cdot[\vec{u_B}(\vec{R_A}+\vec{\delta_i})-\vec{u_A}(\vec{R_A})],
\end{split}
\end{equation}
In the long-wavelength limit, 
\begin{equation}
\vec{u_B}(\vec{R_A}+\vec{\delta_i})-\vec{u_A}(\vec{R_A})=\kappa (\vec{\delta_i} \cdot \nabla)\vec{u},  
\end{equation}
where $\kappa$ is the reduction factor~\cite{suzuura2002phonons}.
Then
\begin{equation}
t_i =t+ \kappa  \frac{\partial \ln{t}}{\partial \ln{a}}\frac{t}{a^2}\vec{\delta_i} \cdot (\vec{\delta_i} \cdot \nabla)\vec{u}.    
\end{equation}
Defining $\beta=-\frac{\partial \ln{t}}{\partial \ln{a}}$, which measures how the bonds respond to being deformed, we have
\begin{equation}
t_i =t- \frac{\kappa \beta t}{a^2}\vec{\delta_i} \cdot (\vec{\delta_i} \cdot \nabla)\vec{u}.   
\end{equation}
Plugging $\vec{\delta}_1=(0,a)$, $\vec{\delta}_2=(-\frac{\sqrt{3}a}{2},-\frac{a}{2})$ and $\vec{\delta}_3=(\frac{\sqrt{3}a}{2},-\frac{a}{2})$ into the above equation, we obtain
\begin{align}
\delta t_1&=-\beta t \kappa u_{yy},\\
\delta t_2&=-\beta t \kappa \left(\frac{3}{4}u_{xx}+\frac{1}{4}u_{yy}+\frac{\sqrt{3}}{2}u_{xy}\right),\\
\delta t_3&=-\beta t \kappa \left(\frac{3}{4}u_{xx}+\frac{1}{4}u_{yy}-\frac{\sqrt{3}}{2}u_{xy}\right),
\end{align}
Then we have $A_x=-\frac{\beta \kappa}{2 e a}(u_{xx}-u_{yy})$ and $A_y=2\frac{\beta \kappa}{2 e a}  u_{xy}$.

%$\mathbf{G}_{1} = \frac{4\pi}{\sqrt{3}a^m}(0,1)$, $\mathbf{G}_{2,3} = \frac{4\pi}{\sqrt{3}a^m}(\mp\sqrt{3}/2,-1/2)$ are the reciprocal lattice vectors.

\section{Nearly flat band in monolayer graphene}
\begin{figure}
\centering
\includegraphics{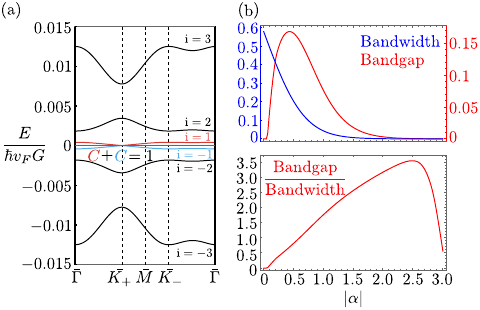}
\caption{Topological flat band in buckled monolayer graphene. (a) Band structure at $\alpha=-2.22817$. Blue line corresponds to band 1 and red line corresponds to band -1. Together, they have valley Chern number 1.(b) Bandwidth(blue line), bandgap(red line in the top panel), ratio between bandgap and bandwidth(red line in the bottom panel) as a function of $|\alpha|$, respectively. Bandgap is the gap between the middle two bands and the band 2, bandwidth is of band -1 or 1. Here band index $i$ means the $i$-th band counting from charge neutrality point, as shown in (a).} 
\label{fig:mono band structure}
\end{figure}

We set ${\delta t_{i}}/{t}= \delta t \sin (\vec{G}_{i} \cdot \vec{r})$, where $\vec{G}_{1} = \frac{4\pi}{3 N_{n} a}(0,1)$, $\vec{G}_{2,3} = \frac{4\pi}{3 N_{n} a}(\mp\sqrt{3}/2,-1/2)$ are reciprocal lattice vectors. The pseudo magnetic field is 
\begin{equation}
\vec{B}(\vec{r})=B_{0}\sum_{i=1}^{3} \cos(\vec{G}_{i} \cdot \vec{r}) \hat{z},
\label{eq:B field}
\end{equation} 
where $B_0=\frac{4\pi t \delta t}{3 e v_{F} N_{n} a}$. We define the normalized $\delta t$ as $\alpha=\frac{\delta t}{\frac{3}{2} a G}$, where $G=\frac{4\pi}{3 N_{n} a}$ is the magnitude of the reciprocal lattice vector. We obtain two nearly flat bands near the charge neutrality point when $\delta t=-0.2$ and $N_n=70$, which corresponds to $\alpha=-2.22817$, as shown in Fig.~\ref{fig:mono band structure}(a). Adding a $m_{0} \sigma_{z}$ term and setting $m_{0}=0.001$, we get two separated flat bands. The valley Chern number is 1 for the bottom flat band and 0 for the top flat band. Before adding $m_{0} \sigma_{z}$ term, the ratio between the bandgap and the bandwidth of band 1/(-1) is around 3, as shown in Fig.~\ref{fig:mono band structure}(b). Here band index $i$ means the $i$-th band counting from charge neutrality point. After adding $m_{0} \sigma_{z}$ term, the ratio between bandgap (the bandgap is between the band 1(-1) and 2(-2) for the top(bottom) band.) and bandwidth of the top(bottom) are the same and around 18.3.

\begin{figure}[b]
\includegraphics{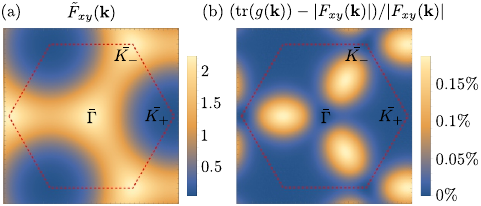} 
\caption{(a) Berry curvature distribution of the topological band in monolayer strained graphene, where $\tilde{F}_{xy}(\mathbf{k})={F_{xy}(\mathbf{k})}/{\bar{F}_{xy}}$. (b) Violation of the trace condition of the topological flat band. Here, we choose $\alpha=-2.22817$ and $m_{0}=0.001$.}
\label{fig:mono trace condi violation}
\end{figure}

In addition to the band dispersion, we further characterize the bands using the Fubini-Study metric $g_{ab}(\mathbf{k})$ and the Berry curvature $F_{xy}(\mathbf{k})$, which are given by:
\begin{equation}
\begin{split}
g_{ab}(\mathbf{k}) &= \Re\left(\eta_{ab}(\mathbf{k})\right), F_{xy}(\mathbf{k}) = -2\Im\left(\eta_{xy}(\mathbf{k})\right),\\
\text{where }\eta_{ab}(\mathbf{k}) &= \frac{\langle \partial_a u_\mathbf{k}| \partial_b u_\mathbf{k}\rangle}{||u_\mathbf{k}||^2}-\frac{\langle \partial_a u_\mathbf{k}|u_\mathbf{k}\rangle\langle u_\mathbf{k}| \partial_b u_\mathbf{k}\rangle}{||u_\mathbf{k}||^4},
\end{split}
\end{equation}
where $u_\mathbf{k}(\mathbf{r})$ is the periodic part of the Bloch function, $\partial_a \equiv \partial_{k_a}$, and $||u_\mathbf{k}||^2 = \langle u_\mathbf{k} |u_\mathbf{k}\rangle$. 
In analogy to the Landau level physics, the ideal candidate for the realization of fractional Chern insulators and other correlated topological states should have an ideal quantum metric, i.e. satisfying the trace condition $\text{tr}(g(\mathbf{k})) =|F_{xy}(\mathbf{k})|$, and a very uniform Berry curvature. To quantify the non-uniformity of the Berry curvature, we plot the Berry curvature distribution of the bottom band in Fig.~\ref{fig:mono trace condi violation}(a). The ratio between the root-mean-square deviation of the Berry curvature and its average value ${\Delta F_{xy}}/\bar{F}_{xy}=0.576179$. We also plot the violation of the trace condition quantified by $[{\text{tr}(g(\mathbf{k})) - |F_{xy}(\mathbf{k})|}]{|F_{xy}(\mathbf{k})|}^{-1}$ in Fig.~\ref{fig:mono trace condi violation}(b). We find that the deviation of $\text{tr}(g(\mathbf{k}))$ from $|F_{xy}(\mathbf{k})|$ is very small ($<0.18\%$) in the entire Brillouin zone for the chosen value of $\alpha = -2.22817$.
\section{Nearly flat bands in bilayer graphene}
% Here we discuss in more detail the effect of strain in bilayer graphene, and the condition under which it can be modeled by an effective gauge field.
Here, we consider Bernal-stacked (AB-stacked) bilayer graphene, which hosts Dirac fermions with winding number 2. The generic valley projected Hamiltonian of the strained bilayer graphene is:

\begin{widetext} 
\begin{equation}
H_{K_{+}}= \frac{3}{2} a t 
\begin{pmatrix}
0 & -2i\partial_{\bar{z}} +e \tilde{A}_{1} & 0 &0 \\
-2i\partial_z+e \tilde{A}_{1}^{*}& 0 & \frac{2\gamma}{3 a t} & 0\\
 0 &  \frac{2\gamma}{3 a t} &0&  -2i\partial_{\bar{z}}+e \tilde{A}_{2}\\
0 & 0 & -2i\partial_z+e \tilde{A}_{2}^{*}& 0\\ 
\end{pmatrix},
\label{eq:bilayer}
\end{equation}
\end{widetext}
where $\tilde{A}_{1}=A_{1x}+i A_{1y}$, $\tilde{A}_{2}=A_{2x}+i A_{2y}$ and $\gamma$ is the interlayer hopping strength. In the following calculation, we set $\gamma=0.1 t$.
We define the symmetric and antisymmetric fields $\tilde{A}^{(+)}=\frac{\tilde{A}_1+\tilde{A}_2}{2}$ and $\tilde{A}^{(-)}=\frac{\tilde{A}_1-\tilde{A}_2}{2}$, respectively. 
In Bernal stacked bilayer graphene, the aligned top layer $A$ sublattice and the bottom layer $B$ sublattice hybridize strongly and are responsible for the formation of bands away from zero energy. Therefore, we can project out these two interlayer aligned sublattices and derive a low-energy Hamitlonian for the unaligned sublattice. Then the 2 $\times$ 2 effective Hamiltonian of~\eqref{eq:bilayer} is:
\begin{equation}
H_\text{eff}= \frac{9 a^2 t^2}{4\gamma}
\begin{pmatrix}
0 & H_{12} \\
H_{12}^{*} & 0\\ 
\end{pmatrix},
\label{eq:bilayereff}
\end{equation} 
where 
\begin{equation}
\begin{split}
    H_{12}&=-\left(-2i\partial_{\bar{z}}+e\tilde{A}^{(+)}+e\tilde{A}^{(-)}\right)\left(-2i\partial_{\bar{z}}+e\tilde{A}^{(+)}-e\tilde{A}^{(-)}\right)\\
    &= 4\partial_{\bar{z}}^2 -e^2\left(\tilde{A}^{(+)}\right)^2+e^2\left(\tilde{A}^{(-)}\right)^2+4ei\tilde{A}^{(+)}\partial_{\bar{z}}\\
    &\phantom{=}+2ei\partial_{\bar{z}}\tilde{A}^{(+)}-2ei\partial_{\bar{z}}\tilde{A}^{(-)}.
\end{split} 
\end{equation}
% \begin{equation}
% \begin{split}
% H_\text{eff} =&\frac{9 a^2 t^2}{4\gamma}\bigg[-(q_{x}^{2}-q_{y}^{2})\sigma_{x}+2 q_{x} q_{y} \sigma_{y}\\
% &\phantom{\frac{9 a^2 t^2}{4\gamma}\bigg[}+e^2\left({(A_{y}^{(+)}}^{2}-{A_{x}^{(+)}}^{2})\sigma_{x}+2A_{x}^{(+)} A_{y}^{(+)}\sigma_{y}\right)\\
% &+\frac{2e}{3 a t}(2(-A_{x}^{(+)}q_{x}+A_{y}^{(+)}q_{y})\sigma_{x}+(A_{x}^{(+)}q_{y}+A_{y}^{(+)}q_{x})\sigma_{y}\\
% &+(\partial_{y}A_{y}^{(+)}-\partial_{x}A_{x}^{(+)})\sigma_{y}-(\partial_{x} A_{y}^{(+)}+\partial_{y} A_{x}^{(+)})\sigma_{x}) \\
% &\phantom{\frac{9 a^2 t^2}{4\gamma}\bigg[}+e^2\left(({A_{x}^{(-)}}^{2}-{A_{y}^{(-)}}^{2})\sigma_{x}-2A_{x}^{(-)} A_{y}^{(-)}\sigma_{y}\right)\\
% &\phantom{\frac{9 a^2 t^2}{4\gamma}\bigg[} +\frac{2e}{3 a t}\left((\partial_{x} A_{y}^{(-)}+\partial_{y} A_{x}^{(-)})\sigma_{x}+(\partial_{x} A_{x}^{(-)}-\partial_{y} A_{y}^{(-)})\sigma_{y}\right)\bigg]
% \end{split} 
% \label{eq:general2by2}
% \end{equation}
It is evident from Eq.~\eqref{eq:bilayereff} that when the strain field is symmetric with $\tilde{A}^{(-)} = 0$ (i.e., the same for both layers), it acts as a gauge field ($-2i\partial_{\bar{z}} \rightarrow -2i\partial_{\bar{z}} + e\tilde{A}^{(+)}$). However, the antisymmetric part of the strain field $\tilde{A}^{(-)}$ cannot be treated as an effective gauge field. Note that the symmetric and antisymmetric components of the strain field $\tilde{A}$ do not mix. In the following, we discuss these two cases separately.

% We consider two cases: (1) two layers have opposite strain field, (2) two layers have the same strain field. 
\subsection{Bilayer graphene with symmetric strain for two layers}
\begin{figure}
\centering
\includegraphics{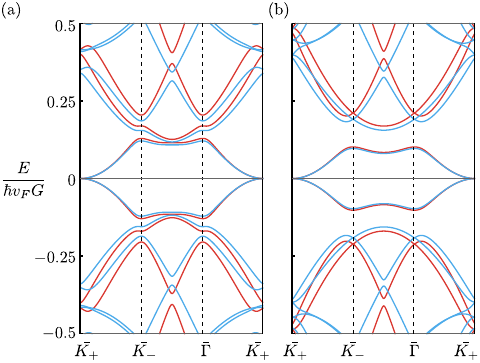} 
\caption{(a) 4 $\times$ 4 (Blue line) and 2 $\times$ 2 (Red line) band structure for symmetric strained bilayer graphene with $\alpha=0.159155$. (b) 4$\times$ 4 (Blue line) and 2$\times$ 2 (Red line) band structure for antisymmetric strained bilayer graphene with $\alpha=-0.159155$.}
\label{fig:42match}
\end{figure}

When the two layers have the same strain, $\tilde{A}^{(-)}=0$. Setting $\tilde{A}^{(+)}=A_{x}+i A_{y}$ and 
%The 4 $\times$ 4 Hamiltonian is:
% \begin{equation}
% \frac{H_{K_{+}}}{\frac{3}{2} a t }=
% \begin{pmatrix}
% 0 & -2i\partial_{\bar{z}} +e \tilde{A}^{(+)} & 0 &0 \\
% -2i\partial_z+e (\tilde{A}^{(+)})^{*}& 0 & \frac{2\gamma}{3 a t} & 0\\
%  0 &  \frac{2\gamma}{3 a t} &0&  -2i\partial_{\bar{z}}+e \tilde{A}^{(+)}\\
% 0 & 0 & -2i\partial_z+e (\tilde{A}^{(+)})^{*}& 0\\ 
% \end{pmatrix},
% \end{equation}
% then the effective 2 $\times$ 2 Hamiltonian becomes
% \begin{equation}
% H_\text{eff}= \frac{9 a^2 t^2}{4\gamma}
% \begin{pmatrix}
% 0 & H_{12} \\
% H_{12}^{*} & 0\\ 
% \end{pmatrix},
% \label{eq:bilayereff}
% \end{equation} 
% where
% \begin{equation}
%     H_{12} = 4\partial_{\bar{z}}^2 -e^2\left(\tilde{A}^{(+)}\right)^2+4ei\tilde{A}^{(+)}\partial_{\bar{z}}+2ei\partial_{\bar{z}}\tilde{A}^{(+)}.
% \end{equation}
\begin{equation}
A_{x}=\frac{-1}{e v_{F}} t \delta t[\sin(\vec{G_{1}} \cdot \vec{r})-\frac{1}{2}\sin(\vec{G_{2}} \cdot \vec{r})-\frac{1}{2}\sin(\vec{G_{3}} \cdot \vec{r})]
\label{eq:Ax}
\end{equation}
\begin{equation}
A_{y}=\frac{-1}{e v_{F}} t \delta t [\frac{\sqrt{3}}{2}\sin(\vec{G_{2}} \cdot \vec{r})-\frac{\sqrt{3}}{2}\sin(\vec{G_{3}} \cdot \vec{r})].
\label{eq:Ay}
\end{equation}
We plot the band structure of the 4 $\times$ 4 Hamiltonian and the 2 $\times$ 2 effective Hamiltonian and find that they agree well near the charge neutrality point when the strength of the strain field is small,as shown for $\alpha=0.159155$ in Fig.~\ref{fig:42match}(a).
Because $ \rho(C_2 T) H_{\text{eff}}^{*}(\mathbf{r},\delta t) \rho(C_2 T)^{\dagger} = H_{\text{eff}}(C_2\mathbf{r},-\delta t)$, the band structure for $\delta t$ and $-\delta t$ are the same. And both $\delta t>0$ and $\delta t<0$ lead to a nonzero bandgap between the middle two bands and the other bands. Before adding any $\sigma_z$ term, the band 2 already has a valley Chern number +2/-2, depending on the sign of $\delta t$. When $\delta t<0$,after adding $\sigma_z$ term, the band -1, 1 and 2 have valley Chern number -2,0,2, respectively. When $\delta t>0$, after adding $\sigma_z$ term, the bands -1,1,2 have valley Chern number 0,2,-2, respectively.
%n is the size of Gtable
\subsection{Bilayer graphene with opposite strain for two layers}
\begin{figure}
\centering
\includegraphics{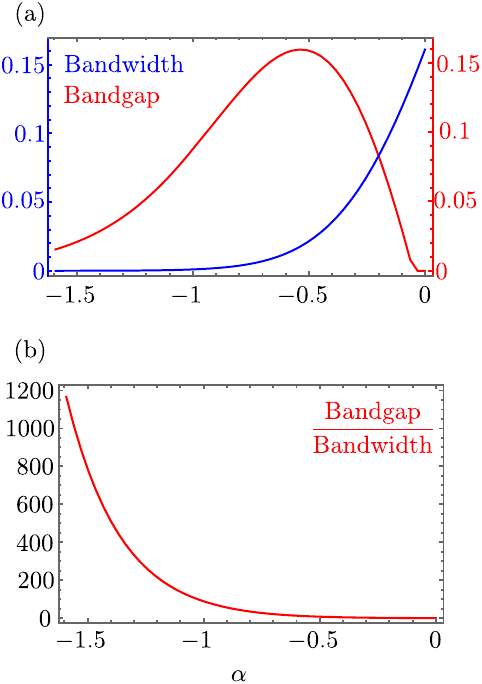} 
\caption{Bandgap and bandwidth plot of antisymmetric strained bilayer graphene. (a) Bandwidth(blue line) of band 1/(-1), bandgap(red line) between band 1(-1) and band 2(-2) and (b) the ratio between bandgap and bandwidth(red line) as a function of $\vert \alpha \vert$. The plot is from the $4\times 4$ Hamiltonian. The normalized $\delta t$ is defined as $\alpha=\frac{\delta t}{\frac{3}{2} a G}$. The energy is shown in units of $\hbar v_{F} G$.}
\label{fig:asy bg/bw}
\end{figure}

%\delta t>0 and \delta t<0 are not symmetric
%\subsubsection{4 $\times$ 4 Hamiltonian}
When the two layers have the opposite strain, we have $\tilde{A}^{(+)}=0$. 
% The 4 $\times$ 4 Hamiltonian is:
% \begin{widetext} 
% \begin{equation}
% H_{K_{+}}=
% \frac{3}{2} a t \begin{pmatrix}
% 0 & -2i\partial_{\bar{z}} +e \tilde{A}^{(-)} & 0 &0 \\
% -2i\partial_z+e (\tilde{A}^{(-)})^{*}& 0 & \frac{2\gamma}{3 a t} & 0\\
%  0 &  \frac{2\gamma}{3 a t} &0&  -2i\partial_{\bar{z}}-e \tilde{A}^{(-)}\\
% 0 & 0 & -2i\partial_z-e (\tilde{A}^{(-)})^{*}& 0\\ 
% \end{pmatrix}.
% \end{equation}
% \end{widetext}
% Consider the valley-projected Hamiltonian for oppositely strained bilayer graphene around $K_{+}$
% \begin{widetext} 
% \begin{equation}
% H_{K_{+}}= \frac{3}{2} a t 
% \begin{pmatrix}
% 0 & q_{x}+i q_{y}+e(A_{x}+i A_{y}) & 0 &0 \\
%  q_{x}-i q_{y}+e(A_{x}-i A_{y})& 0 & \frac{2\gamma}{3 a t} & 0\\
%  0 &  \frac{2\gamma}{3 a t} &0&  q_{x}+i q_{y}-e(A_{x}+i A_{y})\\
% 0 & 0 & q_{x}-i q_{y}-e(A_{x}-i A_{y})& 0\\ 
% \end{pmatrix}
% \end{equation}
% \end{widetext}
We set $\tilde{A}^{(-)}=A_{x}+i A_{y}$, where $A_{x}$ and $A_{y}$ are the same as in Eq.~\eqref{eq:Ax} and  ~\eqref{eq:Ay}. 
% \begin{equation}
% A_{x}=\frac{-1}{e v_{F}} t \delta t[\sin(\vec{G_{1}} \cdot \vec{r})-\frac{1}{2}\sin(\vec{G_{2}} \cdot \vec{r})-\frac{1}{2}\sin(\vec{G_{3}} \cdot \vec{r})]
% \end{equation}
% \begin{equation}
% A_{y}=\frac{-1}{e v_{F}} t \delta t [\frac{\sqrt{3}}{2}\sin(\vec{G_{2}} \cdot \vec{r})-\frac{\sqrt{3}}{2}\sin(\vec{G_{3}} \cdot \vec{r})],
% \end{equation}
Unlike the case of symmetric strain, for antisymmetric strain, positive and negative $\delta t$ lead to different behavior. When $\delta t >0$, the middle two bands are gapless. When $\delta t <0$, the two middle bands are separated from others, and the flatness of the two middle bands increases monotonically as we increase $\vert \alpha \vert$, as shown in Fig.~\ref{fig:asy bg/bw}.

\begin{figure}
\centering
\includegraphics{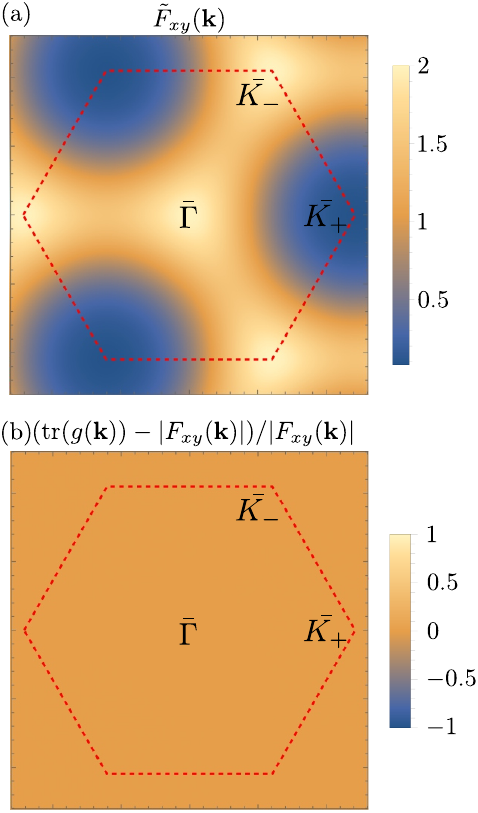} 
\caption{(a) Distribution of the Berry curvature. (b) Violation of the trace condition of the lower topological flat band for antisymmetric strained bilayer graphene at $\alpha=-1.59155$ and $m_0=0.0001$. The plots are from the $4\times 4$ Hamiltonian. $m_0$ is amplitude of $\mathbb{1} \otimes \sigma_z$ term.}
\label{fig:8}
\end{figure}

\begin{figure}
\centering
\includegraphics{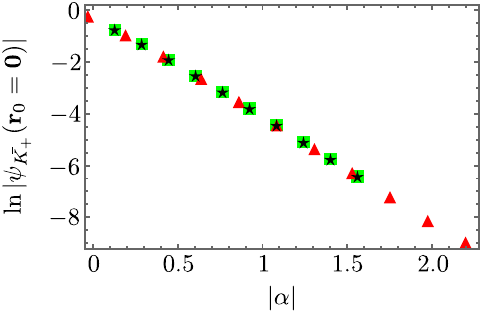} 
\caption{Plot of $\ln{\vert \psi_{\bar{K_{+}}}(\mathbf{r_0}=\mathbf{0}) \vert}$ as a function of $|\alpha|$ for monolayer, bilayer 4 $\times$ 4 Hamiltonian and bilayer 2$\times$2 effective Hamiltonian. Red triangle is for monolayer, Black star is for bilayer 2$\times$2, Green square is for bilayer 4$\times$4.}
\label{fig:7}
\end{figure}

\begin{figure}
\centering
\includegraphics{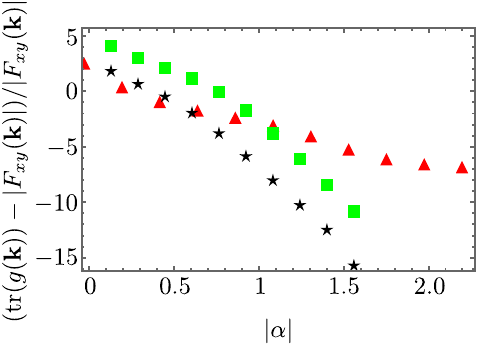} 
\caption{Trace condition violation on a log scale for monolayer and bilayer 4 $\times$ 4 and 2 $\times$ 2 effective Hamiltonian as a function of $|\alpha|$ at a fixed $k$ point. We choose the middle point between $\bar{\Gamma}$ and $\bar{K_{-}}$ for the monolayer case. We choose $\bar{\Gamma}$ for the bilayer case. Red triangle is for monolayer, Black star is for bilayer 2$\times$2, Green square is for bilayer 4$\times$4.}
\label{fig:9}
\end{figure}

We calculate the valley Chern number of the middle two bands after adding $\sigma_z$ term, the top flat band has valley Chern number -1 and the lower one has valley Chern number +1. We plot the Berry curvature distribution and the violation of the trace condition $[{\text{tr}(g(\mathbf{k})) - |F_{xy}(\mathbf{k})|}]{|F_{xy}(\mathbf{k})|}^{-1}$ of the bottom band at $\alpha=-1.59155$. The ratio between bandgap and bandwidth is greater than 1000 after adding $\sigma_{z}$, and we have an ideal quantum geometry, as shown in Fig.~\ref{fig:8}(b). $\vert \psi_{\bar{K_{+}}}(\mathbf{r}_0=\mathbf{0}) \vert$ decrease exponentially as a function of $\vert \alpha \vert$, as shown in Fig.~\ref{fig:7}. When $\vert \alpha \vert$ is large, $\vert \psi_{\bar{K_{+}}}(\mathbf{r_0}=\mathbf{0}) \vert$ is very close to 0, suggesting that the wavefunction of the topological flat band is close to a trial wavefunction $\psi_\mathbf{k}(\mathbf{r})=f_\mathbf{k}(z)\psi_{\bar{K}_{+}}(\mathbf{r})$, where $\psi_{\bar{K}_{+}}(\mathbf{r})$ has a zero at $\mathbf{r}=0$. This trial wavefunction's periodic part is holomophic in $k$, which gives rise to the perfect quantum geometry of the band. As shown in Fig.~\ref{fig:9}, the decay rate of $[{\text{tr}(g(\mathbf{k})) - |F_{xy}(\mathbf{k})|}]{|F_{xy}(\mathbf{k})|}^{-1}$ as a function of $\vert \alpha \vert$ for the bilayer is much faster than that of the monolayer, suggesting that the quantum geometry of the bilayer topological flat band is closer to the ideal quantum geometry than the monolayer topological flat band with the same $\alpha$.

\section{Connection to the Jackiw-Rebbi zero mode}
For Dirac fermions coupled to a gauge field, it was proved that there exist zero modes, whose number is the same as the number of total flux minus one in the system~\cite{PhysRevA.19.2461}. 
Whereas this applies even to the situation where the magnetic field is not uniform in space, it doesn't apply to the case where the net magnetic flux is zero. In strained graphene systems, the net magnetic flux is zero so the calculations of Ref.~\cite{PhysRevA.19.2461} are no longer applicable. 
Instead, the nearly flat band in monolayer graphene system can be understood from the perspective of zero mode of Dirac fermion in 1+1 D with a sign-changing mass term, the Jackiw-Rebbi (JR) zero mode~\cite{PhysRevD.13.3398}. To build intuition, we first show that zero modes in graphene under uniform  magnetic field (zeroth Landau level) can be captured with JR method, and how it can be extended to capture partially flat bands in graphene under a (pseudo)magnetic field that varies only in one spatial direction.
%zeroth Landau level in graphene can be understood in terms of the JR mode. 
Then we show details of JR model in strained monolayer graphene.

For a Dirac Hamiltonian in $1+1$ D, $\mathcal{H} (x)=-i \partial_x \sigma_x+ m(x)\sigma_y$ , with a sign-changing mass term, i.e. $m(x\ge 0)\ge 0$ and $m(x<0)<0$, it supports a massless fermionic mode localized at the domain wall $x=0$. For zeroth Landau level in graphene, we take the gauge $A_y=B_z x$ and $A_x=0$, such that the momentum $k_y$ in the $y$ direction is a good quantum number. For each $k_y$, the Dirac Hamiltonian in one valley $K$ becomes $\mathcal{H}(x)_K=- i \partial_x \sigma_x+ (k_y- B_z x)\sigma_y $ which is exactly the 1+1 D Dirac Hamiltonian with a spatially dependent mass term, $m(x)=k_y- B_z x$. For each $k_y$, there exists a JR zero mode at $x_{JR}=k_y/B_z$. The degeneracy of the zeroth Landau level is equal to the number of allowed $k_y$, which equals $N_y=L_x/x_{JR}=L_x L_y B_z/(2\pi)=\Phi/\Phi_0$, where $L_x$ and $L_y$ are the linear system sizes, and we take $k_y$ to be discretized in a step of $2\pi/L_y$. This reproduces the well-known results that the degeneracy of the zeroth Laudau level is the same as the number of flux in units of flux quantum. In the presence of an electric field $E_y$ along the $y$ direction, $k_y=E_y t$ from the semi-classical equation of motion for electrons. After one period, the position of the zero mode in the $x$ direction is exactly shifted by one period, which corresponds to quantized electrical Hall conductance $\sigma_{xy}=1/2\pi$. Therefore, the zeroth Landau level is topological. The JR model reproduces the well-known facts about the zeroth Landau level of Dirac fermions in $2+1$ D.
 
\begin{figure}[t]
\begin{center}
\begin{minipage}{0.5\textwidth}
\includegraphics[scale=1]{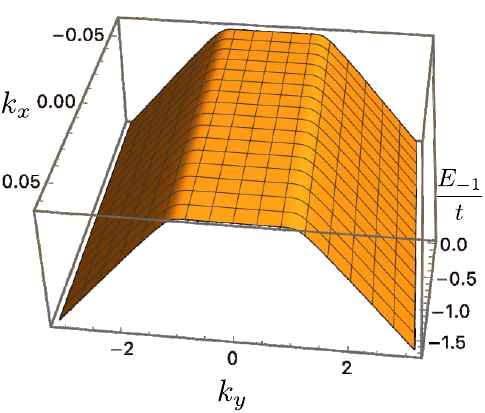}
\end{minipage}\hfill
\end{center}
\caption{Band structure of the band -1 of Hamiltonian of Eq.~\ref{eq:17} as a function of $k_{x}$ and $k_{y}$. Here we set $t=1$, $\sqrt{3}a=1$ and $\delta t=1$.}
\label{fig:jr1}
\end{figure}

% This understanding allows us to choose the proper gauge field by designing the strain field to achieve a nearly flat band. 
However, for a gauge field which corresponds to a uniform magnetic field, the associated strain field diverges at large distances and thus cannot be realized physically. To avoid divergence in the strain field, we choose a sinusoidal vector potential varying only in the $x$ direction, which can be realized using the deformation field $u_y \propto \sin(\frac{2\pi}{N_{n}} \frac{x}{\sqrt{3} a})$ and $u_x=0$.
%\SZL{XH, add here}
Note that there are infinitely many possible strain field configurations that give the same pseudo magnetic field. The Hamiltonian is
\begin{align}
    \mathcal{H}(x, k_y)=v_{F}(-i \partial_{x} \sigma_{x}+(k_{y}+e A_{y}(x)) \sigma_{y})
    \label{eq:17}
\end{align}
where we take $A_{y}(x)=\frac{t}{e v_{F}} \delta t \cos(\frac{2\pi}{N_{n}} \frac{x}{\sqrt{3} a}) $. In the region where $k_y<|\frac{t \delta t}{v_F}|$, there exists a quasi-zero mode localized in the region when $k_y+ e A_y(x)$ changes sign. Here ``quasi-zero" means that these localized modes generally hybridize, which lifts these modes away from zero energy. The degree of hybridization depends on the ratio between the spread of the wave function and the separation of the zeros of $k_y+e A_y(x)$. As $k_y$ approaches $|\frac{t \delta t}{v_F}|$, the hybridization of the localized mode becomes stronger and the energy of the modes starts to deviate from zero energy. As a consequence, the band is nearly flat only in a certain range of $k_y$ as illustrated in Fig.~\ref{fig:jr1}. The partially flat band result is consistent with the tight binding model result in Fig. 3(b) of ~\cite{meng2013strain}. We remark that the partially flat band can also host correlated quantum states, when the Fermi energy is parked in the flat band region.  

Now we are in a position to discuss the periodic strain considered in Sec. III. We take advantage of the gauge redundancy by choosing $A_y=0$ and $A_{x}=\frac{-1}{e v_{F}} t \delta t[ \sin(\vec{G_{1}} \cdot \vec{r})-2\sin(\vec{G_{2}} \cdot \vec{r})-2\sin(\vec{G_{3}} \cdot \vec{r})] $. A strain tensor needs to satisfy a compatibility equation for it to correspond to a single valued displacement field~\cite{Barber}. Using the compatibility equation ($\partial_x^{2}u_{yy}+\partial_y^{2}u_{xx}-2\partial_{xy}^{2}u_{xy}=0$), we find that the strain field corresponds to displacement field $u_x = -\sqrt{3} u_0(\cos (\vec{G_{2}} \cdot \vec{r}-\cos (\vec{G_{3}} \cdot \vec{r})$ and $u_y=u_0 \sum_{i=1}^{3}\cos (\vec{G_{i}} \cdot \vec{r})$, where $u_0=\frac{9 e B_0 N^2 a^3 }{8\pi^{2} \hbar \beta}$. Then $-\partial_y A_x(x,y)=B_z$ is the same pseudo magnetic field as  in Eq.~\eqref{eq:B field}. The Hamiltonian becomes
\begin{equation} 
H=v_{F}(-i \partial_{x} \sigma_{x}-i \partial_{y} \sigma_{y}-eA_{x} \sigma_{x})
\label{eq:jr monolayer}
\end{equation}
Assuming that the variation in the $y$ direction is slow and its amplitude is weak, we use the mode expansion 
\begin{equation}
A_{x,m}(y) \equiv \frac{1}{L_x}\int_0^{L_x} A_x(x,y) \exp(-i m G_x x) dx, 
\label{eq:Am}
\end{equation}
where $G_{x}=\frac{2\pi}{\sqrt{3}N_{n}a}$.
Plug in ~\eqref{eq:Am} and $\psi(x,y)=\sum_{m} e^{i (k_{x}+m G_{x})x} \psi_m(y)$ into the eigenvalue equation of ~\eqref{eq:jr monolayer}, we get
\begin{equation}
[-i\partial_{y}\sigma_{y}+(k_{x}+m G_{x}) \sigma_{x}]\psi_{m}-\sum_{m^{'}}A_{x,m-m^{'}}\sigma_{x}\psi_{m^{'}}=E\psi_{m} 
\end{equation}
For the $A_x(x,y)$ we use, only $A_{x,0}=\frac{-3 B_{0} N_{n} a}{4\pi} \sin(G_{1y} y)$,$A_{x,-1}=\frac{3 B_{0} N_{n} a}{2\pi} \sin(G_{2y} y)$ and $A_{x,1}=\frac{3 B_{0} N_{n} a}{2 \pi} \sin(G_{3y} y)$ are nonzero. 
We consider the zero mode and set $m=0$, then we rewrite the Hamiltonian near $E=0$ as
\begin{equation}
H=-i \partial_{y} \sigma_{y}-M(y) \sigma_{x}  
\end{equation}
where
\begin{equation}
M(y)=k_{x}-A_{x,0}-A_{x,-1}A_{x,1}\left(\frac{1}{k_{x}+G_{x}-A_{x,0}}+\frac{1}{k_{x}-G_{x}-A_{x,0}}\right)
\end{equation}
% If we set 
% \begin{equation}
% A_{x}=-\frac{1}{e v_{F}} t \delta t[ \sin(\vec{G_{1}} \cdot \vec{r})-2\sin(\vec{G_{2}} \cdot \vec{r})-2\sin(\vec{G_{3}} \cdot \vec{r})] 
% \end{equation}
Plug in the value of $A_{x,0}$, $A_{x,-1}$ and $A_{x,1}$, we obtain
\begin{equation}
\begin{split}
&M(y,k_{x},\delta t)=
\frac{3}{2} a t k_{x}+ t \delta t \sin(G_{1y} y)
-4 t^{2} \delta {t}^{2} \sin(G_{2y} y)\sin(G_{3y} y)\\
&\left(\frac{1}{\frac{3}{2}a t(k_{x}+G_{x})+t \delta t \sin(G_{1y} y)}+\frac{1}{\frac{3}{2}a t (k_{x}-G_{x})+t \delta t \sin(G_{1y} y)}\right)
\end{split}
\end{equation}
If we set $t=1$, we can plot $M(y)$ as a function of $y$ for certain $k_x$ and $\delta t$.
\begin{figure}[t]
\centering
\includegraphics[scale=1]{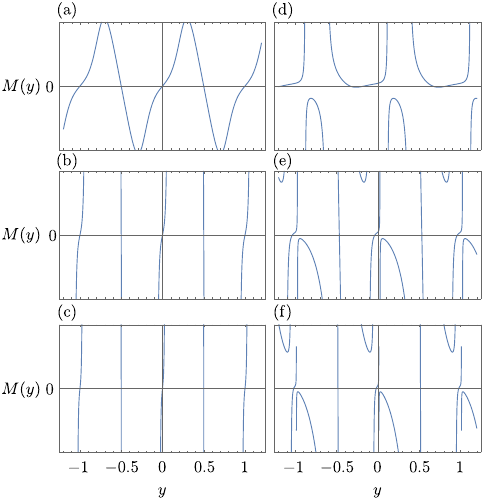}
\caption{Zeros of $M(y)$ of Hamiltonian of Eq. 18 for different $k_x$ and $\delta t$. (a)-(c) $k_x=0$;(d)-(f) $k_{x}=0.05$. (a) and (d) $\delta t=0.05$; (b) and (e) $\delta t=0.2$; (c) and (f) $\delta t=0.4$.} 
\label{fig:JR}
\end{figure}
There are two zeros within one period (see Fig.~\ref{fig:JR}), so there should be two flat bands, which is consistent with Fig.~\ref{fig:mono band structure}. Generally these localized modes hybridize and therefore the band cannot be exactly flat. As shown in Fig.~\ref{fig:JR}, for a specific $k_x$, the slope of $M(y)$ near each zero mode becomes lager as we increase $\delta t$, which means the hybridization of localized mode becomes weaker as we increase $\delta t$ (recall that the larger the slope at the zeros of $M(y)$, the smaller the spread of the wave function of the localized mode.). As a consequence, the middle two bands become flatter as we increase $\delta t$, which qualitatively agree with the bandwidth plot in Fig.~\ref{fig:mono band structure}(b). 
\section{Discussion and conclusion}
First we discuss the existing results of the nearly flat band in graphene and bilayer graphene under a periodic potential. Milovanovi\'c  \emph{et al.} studied band flattening in buckled monolayer graphene numerically by modeling the buckling effect as a pseudo magnetic field. Narrow bands with delocalized electronic dynamics are found~\cite{PhysRevB.102.245427}. Skurativska \emph{et al.} calculated the electronic structure of single layer graphene in the presence of periodic adatoms within the framework of functional density theory, where relatively flat bands with fragile topology are reported~\cite{PhysRevResearch.3.L032003}. The mechanism for the appearance of a nearly flat band in these two works is not clear. In our work, we unravel the underlying mechanism for the flat band through the JR zero mode. In addition, we provide further characterization of the bands using quantum metric. In Ref.~\cite{ghorashi2023topological}, both topological and non-topological flat bands in the bilayer graphene under a superlattice potential are demonstrated. The superlattice potential is introduced through spatially modulated electric field, whereas in our work, the strain modulates the velocity of the Dirac fermion. We would also like to mention that closely related systems of a single Dirac fermion on the surface of a topological insulator in the presence of periodic potential, where flat bands and possible correlated states are studied~\cite{PhysRevX.11.021024,PhysRevB.103.155157}. The flat band for Dirac fermions in the presence of a modulated physical magnetic field with a nonzero net magnetic flux is studied in Ref.~\cite{dong2022dirac}. The appearance of a flat band due to the relative biasing of one sublattice against other sublattices in a bilayer graphene is proposed and verified experimentally in Ref.~\cite{marchenko2018extremely}.

Experimentally, periodic strain can be introduced by placing graphene on top of an array of dielectric nanodots. The anti-symmetric strain for bilayer graphene may be tricky to implement. Generally, one expects that the strain for different layers can be different when the bilayer graphene is placed on a periodic structure. This heterostrain, albeit weak, has been measured experimentally in twisted bilayer graphene devices~\cite{kerelsky2019maximized}. Therefore there exists strain component that is antisymmetric with respect to layer degree of freedom. To maximize the antisymmetric component, one can sandwich the bilayer graphene with periodic structure in both the bottom and top. Although we restrict ourselves to periodic strain with $C_{3v}$ symmetry and strain modulated in one direction, the flat bands can also appear for other periodic strain profiles, as can be inferred from the consideration of the Jackiw-Rebbi zero modes. When the strain can be regarded as a gauge field, because of the gauge redundancy, there exist infinitely many strain profiles that produce the same energy spectrum. Therefore, one may take advantage of this gauge redundancy to find a convenient strain profile for experimental realization. As shown in Fig. \ref{fig:asy bg/bw}, the band flattens as a function of the strain strength. Compared to twisted bilayer graphene, which requires fine tuning of the twist angle to achieve a flat band, strained graphene can achieve a nearly flat band, and the flatness of the band can be improved by continuously tuning the parameters.    

As demonstrated in Sec. IV, the response of the system to a strain field can be very different depending on the winding nubmer of the Dirac fermion. For monolayer graphene with Dirac fermion of winding number 1, strain can be modeled as an effective pseudo gauge field. For high winding numbers, the Dirac fermion can split in the presence of strain, and the pattern of splitting can be controlled by the symmetry of the strain. In this case, strain generally cannot be modeled as pseudo gauge field. For a higher winding number, which can be realized in multilayer graphene, even richer physics can be expected. Similarly, the many-body instability when electron interactions are included is likely distinct for a Dirac fermion with different winding numbers. The response to strain also depends on the position of the Dirac dispersion in the Brillouin zone~\cite{wan2023topological}.

To summarize, we find topological nearly flat bands with a homogeneous distribution of Berry curvature in both monolayer and Bernal-stacked bilayer graphene under periodic strain. Depending on the winding number of the Dirac fermion, the role of strain can either be treated as a pseudo gauge field as in the monolayer graphene, or can be beyond the simple gauge field description as in the case of bilayer graphene. The mechanism of these nearly flat bands can be understood in terms of the Dirac fermion with spatially varying mass terms as in the standard Jackiw-Rebbi model. We further show that the quantum metric of the nearly flat band closely resembles that for Landau levels. These topological flat bands are fertile playground for stabilization of interesting many body quantum states. Similar to other flat band systems~\cite{Ledwith2020Fractional,Repellin2020Chern,Liu2021Gate,Li2021Spontaneous,Xie2021Fractional}, it is easy to expect a valley polarized state and a fractional Chern insulator at frational filling of the valley polarized bands. It is also possible to host other non-abelian topological states due to the nearly ideal quantum metric similar to Landau levels. We leave the interaction effects for future study.

\begin{acknowledgements}
{\it Acknowledgments.---} We thank Chunli Huang, Jacob Pettine, Houtong Chen and Filip Ronning for very helpful discussions. The work at LANL (SZL) was carried out under the auspices of the U.S. DOE NNSA under contract No. 89233218CNA000001 through the LDRD Program, and was supported by the Center for Nonlinear Studies at LANL (XW), and was performed, in part, at the Center for Integrated Nanotechnologies, an Office of Science User Facility operated for the U.S. DOE Office of Science, under user proposals $\#2018BU0010$ and $\#2018BU0083$. This work was supported in part by the Office of Navy Research MURI N00014-20-1-2479 (XW, SS and KS) and Award N00014-21-1-2770 (XW and KS), and by the Gordon and Betty Moore Foundation Award N031710 (KS). 

\end{acknowledgements}

{\it Note added.---} During the preparation of the manuscript, we are aware of the work~\cite{gao2023untwisting} that overlaps in part with the band structure of monolayer graphene under strain in Fig. ~\ref{fig:mono band structure}.
\appendix
\section{Experimental feasibility of nearly flat bands in monolayer graphene}
To obtain the pseudo magnetic field in Eq.~\eqref{eq:B field}, we can put graphene on a substrate which has height profile $h(\mathbf{r})=h_0\sum_{i=1}^{3} \cos(\vec{G}_{i} \cdot \mathbf{r}+\frac{\pi}{4})$ ~\cite{PhysRevLett.128.176406}.
We know 
\begin{equation}
t \delta t=\frac{3 e v_F B_0 N_n a}{4 \pi},
|B_0| \sim (6 \times 10^5 T/\AA^{2}) \times \frac{{h_0}^2}{N_{n}^3}, \text{ and } \alpha=\frac{\delta t}{\frac{3}{2} a G}.  
\end{equation}
Plug in the value of $e$, $v_{F}=\frac{3}{2} a t$, and $a=1.42 \AA$ ($a$ is the bond length of graphene), we obtain
\begin{equation}
    |\alpha| \sim \frac{10}{3 \pi N_n} (h_0/\AA)^2.
\end{equation}
To realize the nearly flat bands at $2.22817 \leq |\alpha| \leq 2.5$ in experiment, we can either utilize nanosphere~\cite{zhang2018strain} or nanopillars ~\cite{jiang2017visualizing} to achieve the height profile. For nanospheres, we can use 20 nm diameter nanospheres to achieve a 20 nm superlattice with a period of $N_n=80$. From Fig.~4e in ~\cite{zhang2018strain}, we get an estimate of $h_0 \sim 17.5 \AA$ for 20 nm diameter nanoparticle case, which corresponds to $|\alpha|=4.06177$. To exactly access  $|\alpha| \sim 2.5$, we just need to use nanospheres with a diameter around 15 nm. For nanopillars, we can arrange triangular nanopillars with period 100 nm and height 2.9 nm on a triangular lattice~\cite{PhysRevLett.128.176406}.

\bibliographystyle{apsrev4-1}
\bibliography{references}

\end{document}